\documentclass[%
 reprint,
prb,
floatfix,
]{revtex4-2}

\usepackage{amsmath,amssymb}
\usepackage{graphicx}
\usepackage{dcolumn}
\usepackage{bm}
\usepackage{xcolor}
\usepackage{siunitx}
\usepackage{hyperref}

\begin{document}

\preprint{APS/123-QED}

\title{Electrical Control of Two-Dimensional Electron-Hole Fluids \\ in the Quantum Hall Regime}
\author{Bo Zou}
\affiliation{Department of Physics, University of Texas at Austin, Austin, TX 78712}
\author{Yongxin Zeng}
\affiliation{Department of Physics, University of Texas at Austin, Austin, TX 78712}
\author{Artem Strashko}\thanks{Current affiliation: Zapata Computing Inc.,100 Federal Street, Boston, MA 02110, USA.}
\affiliation{Center for Computational Quantum Physics, Flatiron Institute, 162 5th Avenue, New York, NY 10010, USA}
\author{Allan H. MacDonald}
\affiliation{Department of Physics, University of Texas at Austin, Austin, TX 78712}
\date{\today}

\begin{abstract}
We study the influence of quantizing perpendicular magnetic fields on the ground state of a bilayer with electron and hole 
fluids separated by an opaque tunnel barrier.  In the absence of a field, the ground state at low carrier densities 
is a condensate of s-wave excitons that has spontaneous interlayer phase coherence. We find that a series of phase transitions emerge at strong perpendicular fields between condensed states and incompressible incoherent states with full electron and hole
Landau levels. When the electron and hole densities are unequal, condensation can occur in higher angular momentum
electron-hole pair states and, at weak fields, break rotational symmetry. We explain how this physics is expressed in dual-gate phase diagrams, and predict transport and capacitively-probed thermodynamic signatures that distinguish different states.
\end{abstract}

\maketitle



\section{Introduction}

Recent progress \cite{wang2019evidence, ma2021strongly,gu2022dipolar, qi2023thermodynamic} in the field of two-dimensional 
materials has made it possible to prepare electron-hole fluids in which the electrons and holes are 
isolated in separate layers and have densities that can be tuned independently.
In the absence of a 
magnetic field, the ground state of neutral electron-hole fluids is 
expected \cite{zhu1995exciton,butov2004condensation,fogler2014high,wu2015theory, varley2016structure,zeng2020electrically} to be an 
exciton-condensate at low carrier densities, and a Fermi liquid at high carrier densities.  
The condensation in electron-hole bilayers, is spatially indirect and implies spontaneous interlayer phase coherence.
The most spectacular properties of bilayer exciton condensates are electrical quantities
that can be measured \cite{tiemann2008exciton, eisenstein2014exciton} when the layers are separately contacted, including
counter-flow superfluidity and giant electrical drag \cite{liu2017quantum, li2017excitonic}.

Spatially indirect exciton condensates were first observed in the quantum
Hall regime and in systems with conduction-band electrons present at the Fermi level in both layers \cite{eisenstein2004bose,eisenstein2014exciton}. 
Spontaneous interlayer phase coherence is then expected to appear 
only in the strong magnetic field quantum Hall regime.  Although the properties of electron-hole bilayers,
in which the carriers in one of the two layers are valence band holes, 
are very similar to those of electron-electron bilayers in the quantum Hall regime,
the two systems behave very differently at weaker magnetic fields.  
In particular, condensation is expected to survive all the 
way to zero magnetic field in the electron-hole case, at least when the system is close to overall neutrality.  
In this article, we analyze, through Hartree-Fock mean-field approach, the crossover between large quantizing magnetic fields and weak magnetic fields in the 
electron-hole case.  We find that exciton-condensate states appear over the full range of magnetic fields,
but that they are interrupted by incoherent states when at least one species is at an integer Landau-level filling 
factor $\nu$.  When the electron and hole densities are unequal, we find that electron-hole pairing at strong fields 
can occur in non-zero angular momentum channels and that states with broken rotational symmetry in the electron-hole pair amplitude  
(nematic exciton insulators) are common at weaker magnetic fields.  

Our paper is organized as follows.  In Section II, we formulate the electrostatics of a prototype of dual-gate electron-hole devices in a way that allows strong electron-hole correlations, essential for the physics 
of interest here, to be conveniently incorporated.  In Section III, 
we explain the mean-field theory that we use to approximate electron-hole many-body states in a strong magnetic field.
Our main results are summarized and discussed in Section IV and Section V. 

\section{Model and electrostatic description} \label{sec:model}

\begin{figure}
\centering

\includegraphics[width=\linewidth]{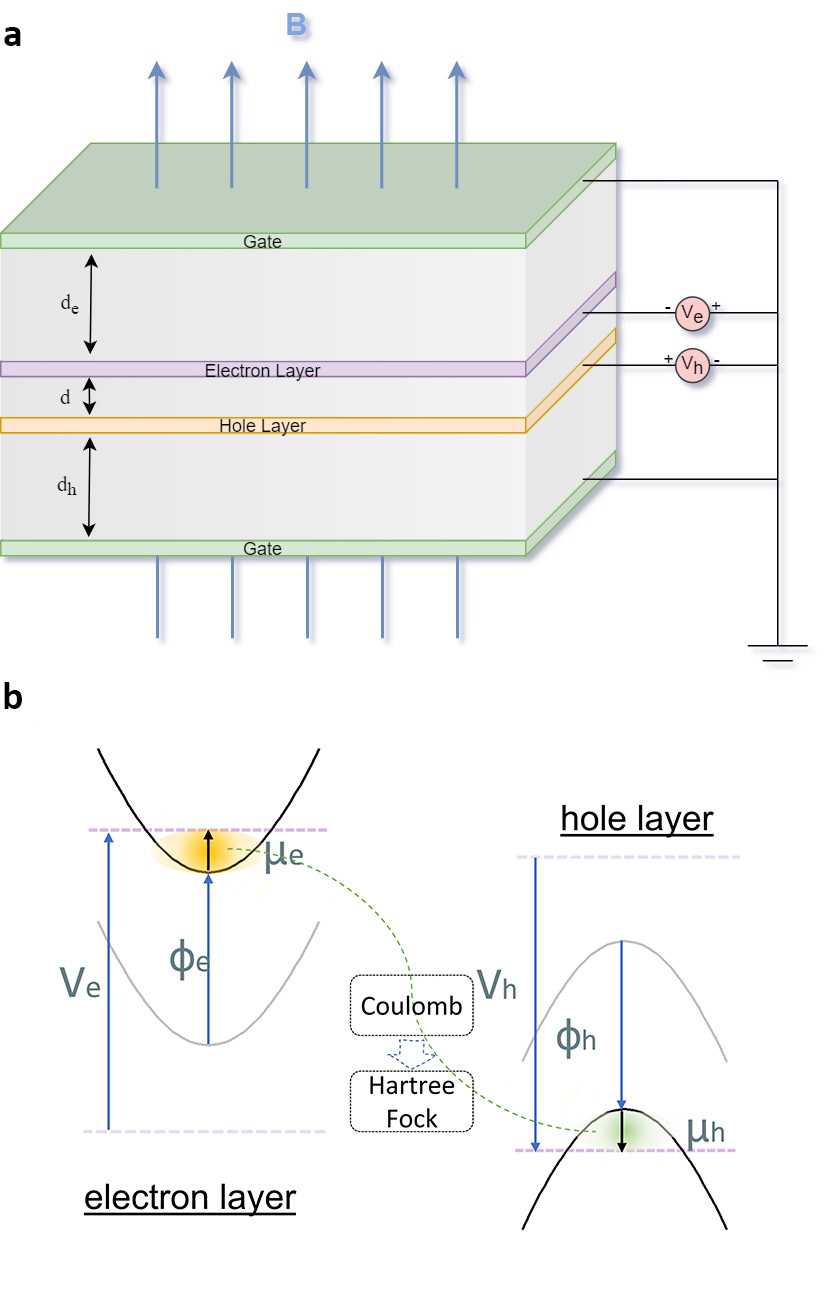}

\caption{(a) Dual-gate semiconductor bilayer device in a strong magnetic field. 
(b) The energy bands and chemical potentials in a quasi-equilibrium state,
with a bias-voltage controlled difference between electron and hole layer 
chemical potentials. The conduction and valence energy bands are shifted by electrostatic potentials. Assuming that the two layers are electrically isolated, the chemical potential difference is equal to the difference between electron and hole layer bias voltages.  The chemical potentials in each layer depend on carrier-density both through band-filling 
effects (shaded yellow for electrons and green for holes in the figure) 
and interaction effects discussed in the main text. 
The difference between the chemical potentials in the two layers 
acts as a chemical potential for excitons.} \label{fig:system}
\end{figure}

We study dual-gated electron-hole bilayer systems like the one illustrated 
schematically in Fig.~\ref{fig:system}(a) in a strong perpendicular magnetic field.  We have in mind in particular transition metal dichalcogenide (TMD) 
single-layer two-dimensional semiconductors because this platform allows 
for good electrical isolation between layers that are close enough to maintain strong electron-hole interactions.
In Fig.~\ref{fig:system} carrier densities are controlled by applying voltages $-V_e$ and $V_h$ between gates (held at ground) and the electron and hole layers.
When the gate voltages are tuned to the proper regime (see below), free carriers are injected into the 
layers to form an electron-hole
fluid with electron density $n$ in one layer and hole density $p$ in the other.
(The valence band of electron layer is assumed to be full and the conduction band of the hole layer to be empty, and the signs of $V_e$ and $V_h$ are chosen so that positive voltages induce carriers.)
Critically, the two layers are assumed to be electrically isolated so that the carrier chemical potentials 
in each layer is fixed by its electrical contact.  Because the many-body physics of electron-hole bilayers is  
most conveniently calculated in models that set the electrical potential and the energies of the band extrema to
zero in each layer, it is necessary to take some care in analyzing the relationship between the experimental control parameters,
the gate voltages, to the many-body chemical potentials calculating in interacting electron-hole models.
This analysis has been undertaken in Ref.~\cite{zeng2020electrically}, 
and is briefly repeated below in order to establish 
critical notations.  The analysis is readily generalized to allow for a chemical potential
difference between the two gate layers.

The geometry of our model system implies the following electrostatic relations between the
electron potential energies ($\phi_e$ and $-\phi_h$ for the electron and hole layers) and the carrier densities ($n,p$) of the two layers (in cgs units):
\begin{equation} \label{eq:electrostatics}
    \begin{aligned}
        \phi_e &= \frac{4\pi e^2 d_e}{\epsilon} \frac{(d_h+d)n-d_h p}{d_e+d_h+d},\\
        \phi_h &= \frac{4\pi e^2 d_h}{\epsilon} \frac{(d_e+d)p-d_e n}{d_e+d_h+d}.
    \end{aligned}
\end{equation}
Here $\epsilon$ is the dielectric constant of the tunnel barriers separating the active layers and the gates, 
$d_{e/h}$ is the vertical distance between the electron/hole layer and the top/bottom gate, 
and $d$ is the interlayer distance, which is typically much smaller than $d_e$ and $d_h$.
(For convenience we define the voltages $V_{e/h}$ as quantities with
dimensions of energy by absorbing a factor of electron charge $e$.) In Eqs.~\eqref{eq:electrostatics} we have assumed that both layers have uniform charge densities. In this article we limit our attention to translationally invariant states.

As shown in Fig.~\ref{fig:system}(b), the voltages shift the chemical potential and gather carriers.  These in turn induce potential energies that shift the band extrema.
The equilibrium state of the bilayer is one in which electrons are in equilibrium with the contact to the 
electron layer and holes are in equilibrium with the contact to the hole layer:     
\begin{equation}
    \begin{aligned}
        V_e - \phi_{e\,}(n,p) &=\ \ \epsilon_{c\,}  + \mu_e(n,p), \\
        V_h - \phi_h(n,p) &= -\epsilon_{v}  + \mu_h(n,p), \\
    \end{aligned}
    \label{eq:equilibrium}
\end{equation}
where $\mu_e$ and $\mu_h$ are the electron and hole many-body chemical potentials
calculated in a model with neutralizing charge backgrounds in each layer, and
$\epsilon_{c}$ ($\epsilon_{v}$) are the energies of 
the conduction and valence band extrema in the absence of carriers. 
Given results for $\mu_e(n,p)$ and $\mu_h(n,p)$ from many-body theory, 
the values of $n$ and $p$ for a given set of layer voltages
are determined by satisfying Eqs.~\ref{eq:equilibrium}.

When a magnetic field $B$ is applied perpendicular to the two-dimensional plane, 
the dispersive conduction and valence bands are replaced by two sets of Landau levels with energies  
$\{\epsilon_{n,c} = \epsilon_c + (n+\frac{1}{2})\hbar\omega_c\}$ and 
$\{\epsilon_{n,v} = \epsilon_v - (n+\frac{1}{2})\hbar\omega_c\}$.
For simplicity we assume that the conduction band electrons and valence band holes have the same effective mass $m^*$ 
and thus same cyclotron frequency $\omega_c = eB/m^* c$. 
We also neglect the roles of the 
electron and hole spin degrees of freedom. 
Neither simplification is important for our 
principle conclusions, and our theory can easily be generalized to describe
the experimental properties of particular devices. 
The density of carriers in a single Landau level is $(2\pi l^2)^{-1}$ where 
$l = ({\hbar c}/{eB})^{1/2}$ is the magnetic length.
Below, we express carrier densities in terms of Landau level filling factors $\nu_e=nS/g$ and $\nu_h=pS/g$, 
where $S$ is the area of the sample and $g=S/2\pi l^2$ is the degeneracy of the Landau levels.  The total charge filling factor is $\nu_c=\nu_e-\nu_h$.

\section{Hartree-Fock Mean field theory} \label{sec:method}

We apply Hartree-Fock mean-field theory to a system with 
two sets of Landau levels, conduction band Landau levels
in the electron layer and valence band Landau levels in the 
hole layer.  Hartree-Fock theory provides an accurate description of 
electron-hole pair condensates in both weak and strong coupling limits,
just as BCS theory does for electron-electron pairs.  
The condensate solutions we seek have the 
gate-voltage-dependent exciton chemical potential equal to 
$V_{e} + V_{h}$, and can be mapped to equilibrium exciton-condensates
with chemical potential equal to zero by reducing the effective band 
gap by $V_{e}+V_{h}$.  Making this choice and taking the overall zero of energy at the 
mid-point of the effective gap leads to the interacting Hamiltonian 
\begin{equation}
\label{eq:Hamiltonian}
    \begin{aligned}
        &H\ =\ \sum_{\tau,n,X} \epsilon_{n,\tau}c^\dagger_{\tau,n,X}c_{\tau,n,X}\\
        &+\frac{1}{2S}\sum_{\tau,\tau^\prime}\sum_\textbf{q}\sum_{\substack{n_1,n_2,n_3,n_4\\X_1,X_2,X_3,X_4}}V_{\tau\tau^\prime}(\textbf{q})\\
        &\left< \tau\,n_1 X_1 \middle| e^{-i\textbf{q}\cdot\textbf{r}} \middle| \tau\,n_4 X_4 \right> \left< \tau^\prime\,n_2 X_2 \middle| e^{i\textbf{q}\cdot\textbf{r}} \middle| \tau^\prime\,n_3 X_3 \right>\\
        &\ c^\dagger_{\tau,n_1,X_1}c^\dagger_{\tau^\prime,n_2,X_2}c_{\tau^\prime,n_3,X_3}c_{\tau,n_4,X_4}\ ,
    \end{aligned}
\end{equation}
where $\tau=c,v$ labels bands (layers), $n$ labels Landau levels, and $X$ is a Landau 
guiding center label for degenerate states in the same Landau level. 
In Eq.~\ref{eq:Hamiltonian}, $V_{\tau\tau}(\textbf{q})=\frac{e^2}{\epsilon}\frac{2\pi}{q}$ is the Coulomb interaction within the same layer, 
and interactions between layers have
an extra $e^{-qd}$ factor.
The energy difference between the bottom conduction
and top valence Landau levels is $ \epsilon_{0,c}-\epsilon_{0,v} = \hbar\omega_c
+\epsilon_{c}-\epsilon_{v}$.    
When the bilayer is charged, the Hartree mean fields generated by this Hamiltonian diverge.
These electrostatic potentials should in that case be replaced by those 
discussed in the previous section, which take account of the role of the 
gate electrodes in the electrostatics.  
This replacement, generalizes the conventional neutralizing background {\it ansatz} of 
electron gas theory, and is equivalent to a neutralizing background in the balanced and 
distant gate limit $d_e = d_h \gg d$,
which we have assumed in the specific results results presented in the following sections.

The parameters to specify the system are: 1) magnetic field $B$ which determines the 
Landau level spacing $\hbar\omega_c$ and the Landau level degeneracy $g$; 
2) the effective energy gap $\Delta E = \epsilon_{0,c}-\epsilon_{0,v} - \hbar\omega_c$; 
3) the charge filling factor $\nu_c$; 
and 4) $d$, $d_h$ and $d_e$ that describe the geometry of gates and layers.
In Section \ref{sec:result} we explain how these parameters influence the behavior of the system.

\begin{figure}
\centering
\includegraphics[width=\linewidth]{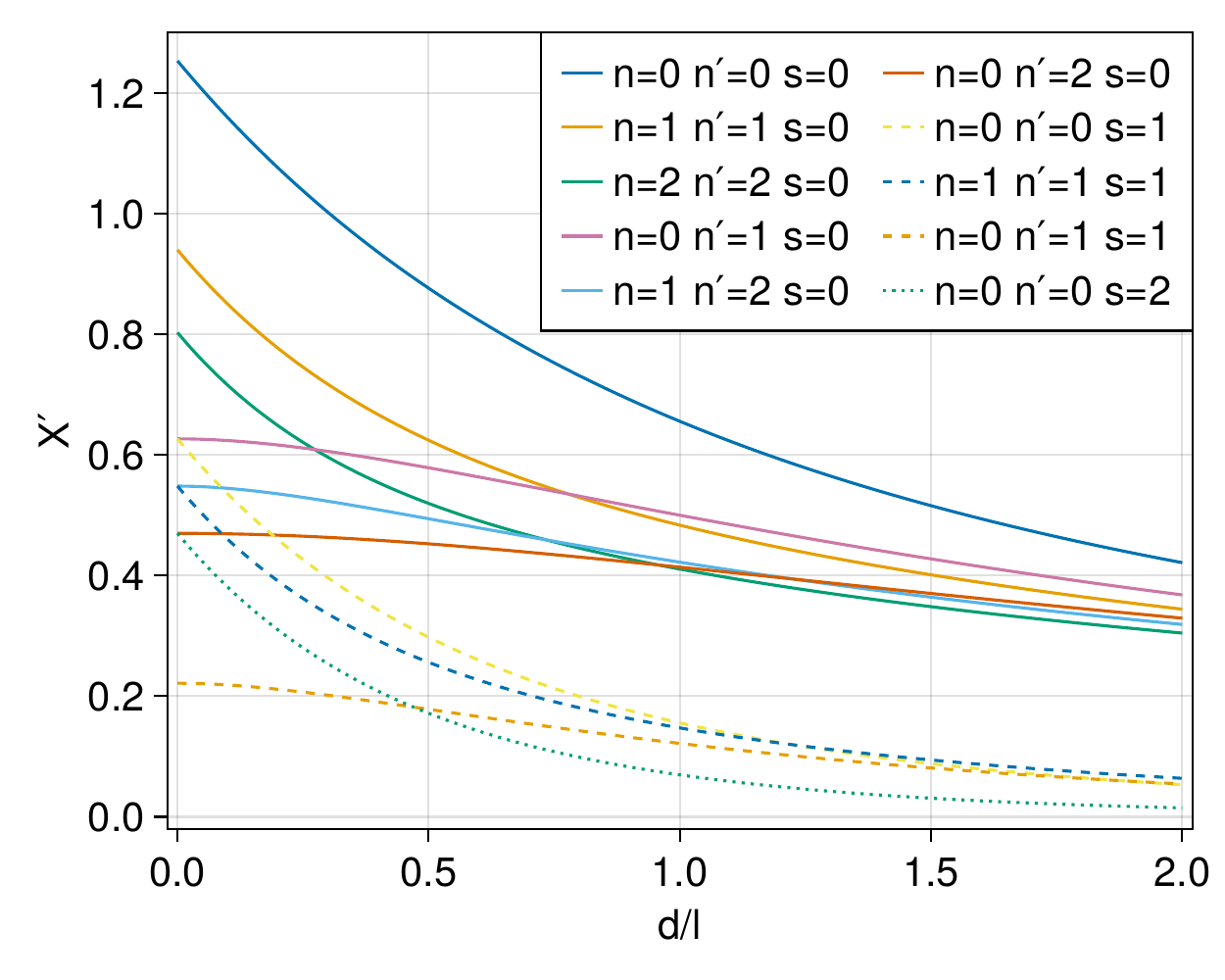}
\caption{Inter-layer exchange integrals $X'$ defined in the text
{\it vs.} layer separation $d$ in units of magnetic length $l$.  The intra-layer exchange integrals $X=X'(d=0)$.  The $(nn';m'm')$ exchange integral specifies the $(n,n')$ matrix-element of the mean-field self-energy operator contributed by the $(m,m')$ density-matrix element, and is non-zero only when $s=m'-n=m-n'$. Exchange is weaker when $s$ is non-zero;
the solid lines are for $s=0$, the dashed lines are for $s=1$,
and the dotted lines for $s=2$.}
\label{fig:X-d}
\end{figure}

\begin{figure*}
\centering

\includegraphics[width=\linewidth]{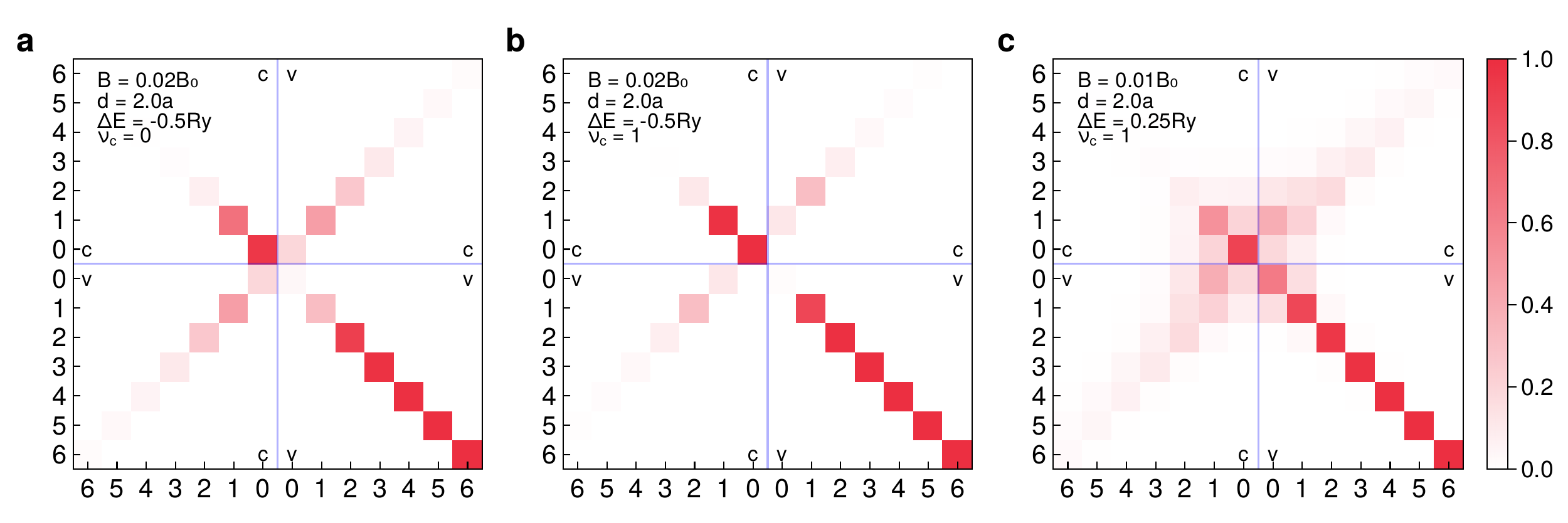}

\caption{Density matrices of three typical XI solutions. Each block plots the magnitude of the
density matrix element for the specified layer and Landau level indices on the color scale at right.
The diagonal elements show the conduction and valence band fillings of the Landau levels.
We have assumed that only the conduction (c) band is relevant in one layer and 
only the valence (v) band is relevant in the other layer.  
The magnetic field $B$, layer separation $d$, effective gap $\Delta E$,
and charge filling factor $\nu_c$ are specified in the upper-left corners of the figures. (a) $\Delta n$=0: only conduction and valence levels with index difference 0 cohere, corresponding to an s-wave condensate; (b) $\Delta n$=1: only conduction and valence levels with index difference 1 cohere corresponding to a $p$-wave condensate; (c) at weak magnetic fields solutions 
appear with non-zero density matrix values for more than one $\Delta n$, implying that the electron-hole pair 
amplitude is not rotationally invariant.} \label{fig:density}
\end{figure*}

Replacing the Hartree terms and adding the exchange terms, we obtain the mean-field Hamiltonian 
\begin{equation}
\begin{aligned}
    H^{(\text{HF})}=\ g \sum_{n}  \Big[(\epsilon_{n,c}&+\phi_{e}) \rho_{\substack{n n\\ c\,c}} + (\epsilon_{n,v}-\phi_{h}) \rho_{\substack{n n\\ v v}} \Big]\\
    &+ g \sum_{n n^\prime\tau \tau^\prime} U(n,n^\prime;\tau,\tau^\prime) \rho_{\substack{n n^\prime\\\tau \tau^\prime}}\ ,
\end{aligned}
\end{equation}
where
\begin{equation}
\label{eq:exchange}
\begin{aligned}
    U(n,n^\prime;\tau,\tau) &= - W_0 \sum_{m,m^\prime}  X_{n n^\prime; m m^\prime} \Big< \rho_{\substack{m m^\prime\\ \tau \tau}} \Big>_r, \\
    U(n,n^\prime;\tau,\overline{\tau}) &= - W_0 \sum_{m,m^\prime}  X^\prime_{n n^\prime; m m^\prime} \Big< \rho_{\substack{m m^\prime\\ \overline{\tau} \tau}} \Big>_r \,,
\end{aligned}
\end{equation}
$W_0 = e^2/\epsilon l$ is the natural interaction energy scale,
and the intralayer ($X$) and interlayer ($X'$) exchange integrals are given explicitly below.  The guiding-center independent density matrices of the spatially uniform 
states we seek is  
\begin{equation}
    \rho_{\substack{n n^\prime\\\tau \tau^\prime}} = g^{-1}\sum_{X} c^\dagger_{\tau,n,X}c_{\tau^\prime,n^\prime,X}\ ,
\end{equation} 
and the electron and hole filling factors are 
\begin{equation}
    \nu_e=\sum_{n} \Big< \rho_{\substack{n n\\c\,c}} \Big> \ ; \ \nu_h=\sum_{n} \left( 1- \Big< \rho_{\substack{n n\\v v}} \Big> \right)\ .
\end{equation}
Note that these solutions we seek correspond to a condensate of electron-hole pairs that have total momentum zero.
We will find that when the total charge density of electrons and holes is non-zero, we sometimes obtain solutions that 
break rotational symmetry.  Under these circumstances the ground state 
condensate in the absence of a magnetic field has \cite{strashko2020crescent} a finite momentum. 
We shall, nevertheless, leave the study of this possibility in the strong field limit to future work.  

\begin{figure*}
\centering
\includegraphics[width=2\columnwidth]{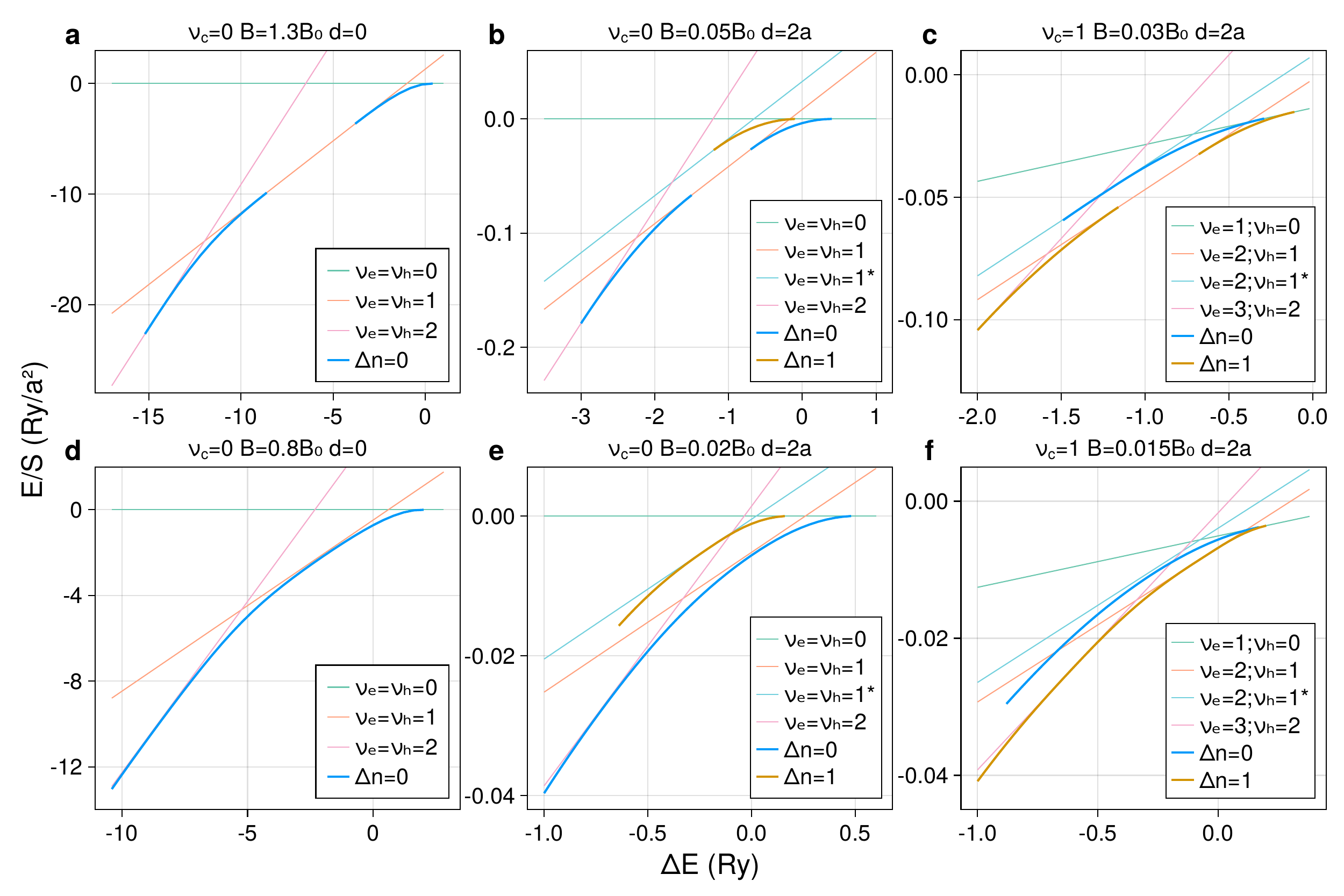}
\caption{Energy per area {\it vs.} effective energy gap $\Delta E$ for 
ground and metastable excited states: (a) charge filling factor $\nu_c$=0, 
layer separation $d$=0 and magnetic field $B$=1.3$B_0$. 
(b) $\nu_c$=0, $d$=2$a$ and $B$=0.05$B_0$. (c) $\nu_c$=1, $d$=2$a$ and $B$=0.03$B_0$. (d) $\nu_c$=0, $d$=0 and $B$=0.8$B_0$. (e) $\nu_c$=0, $d$=2$a$ and $B$=0.02$B_0$. (f) $\nu_c$=1, $d$=2$a$ and $B$=0.015$B_0$. 
The slopes of the energy-$\Delta E$ curves
do not change within the stability ranges of the QHI states
within which $\nu_{e}$ and $\nu_{h}$ are 
fixed integers, specified by the colors indicated in the insets. 
Segments of these strait lines claim the ground state where QHI states are stable and are 
evident in the plots.
The slopes of the electron-hole pair states (XI) curves do vary continuously in their stability ranges,
highlighted by using bold curved lines.
The figure shows that the ground state exciton density $n_{ex}=min\{n, p\}$ increases 
monotonically with decreasing $\Delta E$. 
In neutral systems the interlayer coherence, when it occurs, is always between electron and hole
Landau levels with identical indices ($\Delta n=0$). 
For charge filling factors that are non-zero, states with an electron-hole Landau level index difference $\Delta n \ne 0$ can be ground states.
(Blue and brown curves plot the energies of $\Delta n=0$ and $\Delta n=1$ XI solutions.) 
An asterisk after the electron and hole filling factors in a QHI state legend implies
that the electrons of that state are not filled in order of energy. 
For TMDs the Bohr radius of excitons $a$ is about 1.3nm, the Rydberg energy is about 0.11eV, and $B_0$ is about 2.4kT. 
In comparison these numbers for GaAs are 12nm, 4.6meV, and 27.9T, respectively.
} \label{fig:E-gap}
\end{figure*}

In Eq.~\ref{eq:exchange}, the subscript $r$ implies that the 
density-matrix is to be calculated relative to
$\delta_{mm^\prime}\delta_{\tau\tau^\prime}\delta_{\tau v}$, the density matrix of a
system with a filled valence band and an empty conduction band - whose interacting 
system mean-field is conventionally included in the single-particle 
Hamiltonian.  

The exchange integrals
\begin{equation}
\label{eq:exchangeintegral}
\begin{aligned}
    X_{n n^\prime;m m^\prime}  =&  \left( \frac{n!\ n^\prime!}{m!\,m^\prime!} \right)^\frac{1}{2} \int_0^{\infty}{d(kl)} \ e^{-\frac{1}{2}k^2l^2 } \\
    & \,\times \left( \frac{k^2l^2}{2} \right)^s L_n^{(s)} \left( \frac{k^2l^2}{2} \right) L_{n^\prime}^{(s)} \left( \frac{k^2l^2}{2} \right),\\
    X^\prime_{n n^\prime; m m^\prime} =&  \left( \frac{n!\ n^\prime!}{m!\,m^\prime!} \right)^\frac{1}{2} \int_0^{\infty}{d(kl)} \ e^{-\frac{1}{2}k^2l^2 -kd } \\
    & \,\times \left( \frac{k^2l^2}{2} \right)^s L_n^{(s)} \left( \frac{k^2l^2}{2} \right) L_{n^\prime}^{(s)} \left( \frac{k^2l^2}{2} \right),
\end{aligned}
\end{equation}
vanish unless $s \equiv m^\prime -n=m-n^\prime $.
In Eq.~\ref{eq:exchangeintegral} $s$ is assumed to be 
non-negative and $L_n^{(s)}(x)$ is a generalized Laguerre polynomial.
(The $s\le0$ integrals are given by $X^{(\prime)}_{n n^\prime; m m^\prime}$=$X^{(\prime)}_{m^\prime m; n^\prime n}$.)
Fig.~\ref{fig:X-d} plots some of the $X'$ integrals ${\it vs.}$ 
electron-hole layer separation.  
We see that the exchange strength is weaker for higher Landau level indices
$n$ or $n^\prime$, and for larger $s$.  The inter-layer coefficients decrease when 
layer separation $d$ is increased.

\begin{figure*}
\centering
\includegraphics[width=\linewidth]{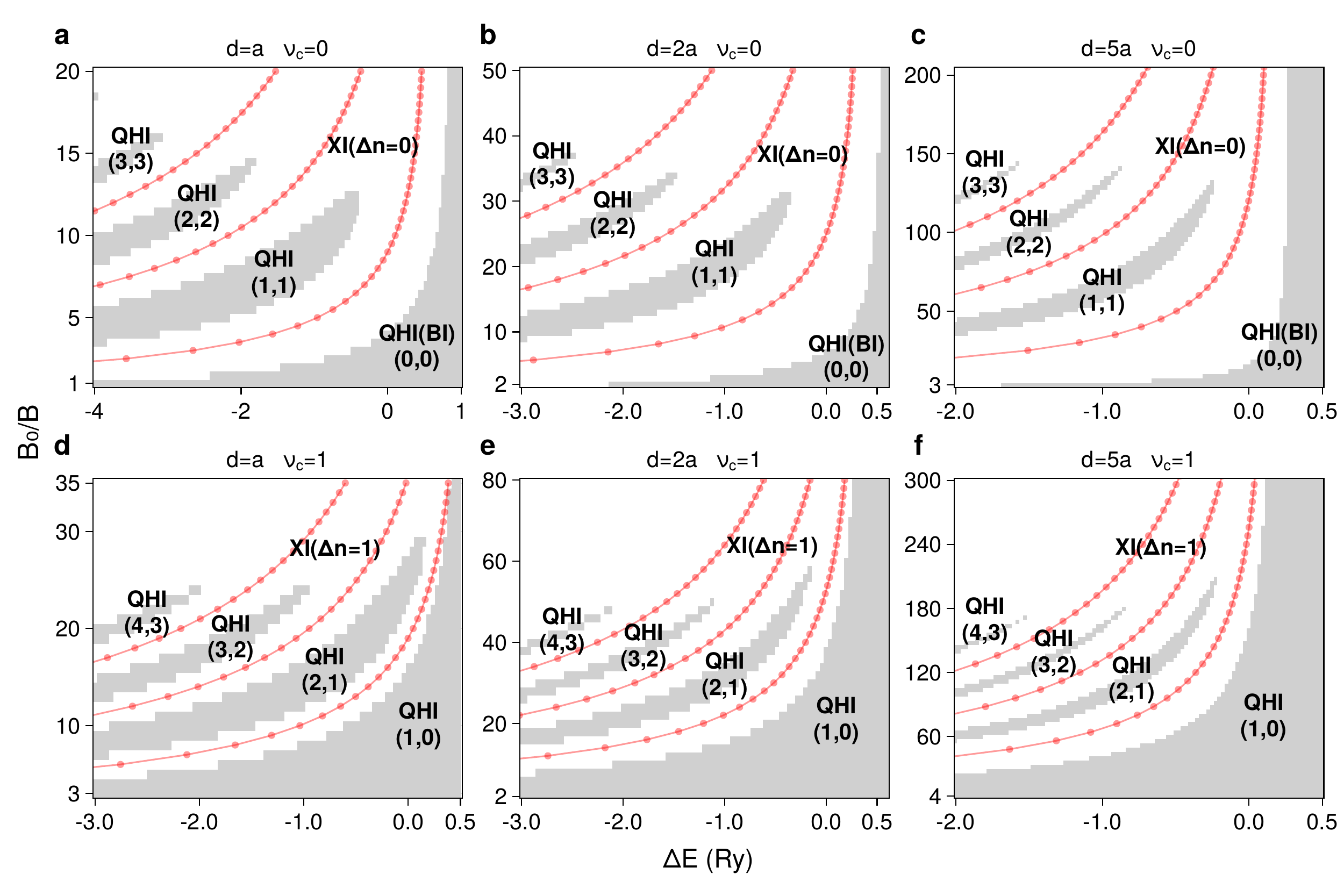}
\caption{Phase diagram {\it vs.} effective gap and dimensionless inverse magnetic field $B_0/B$
where $B_0$ is the magnetic field scale at which the flux through area $a^2$
(where $a$ is the Bohr radius) is a flux quantum $\Phi_0=hc/e$.  For WSe$_2$ parameters,
$B_0$ $\approx 2400$ Tesla, whereas for GaAs $B_0$ $\approx$ 28 Tesla. 
The grey regions are quantum Hall insulators (QHI) with the filling factor of two layers 
($\nu_e$,$\nu_h$) specified. The white regions indicate excitonic insulator (XI) states with 
spontaneous interlayer phase coherence.  The phase diagrams are constructed by dividing
the parameter space into rectangular pixels and performing self-consistent calculations at the center point of each pixel:
(a) layer separation $d=a$ and equal electron and hole Landau level filling factors, (b) and (c) 
like (a) but with layer separations $d=2a$ and $d=5a$,
(d) like (a) but with electron and hole filling factors different by $1$, and (e) and (f)
respectively $d=2a$ and $d=5a$ with filling factor difference $1$.
Pairing is weaker at realistic layer separation - in the $d=2a$ to $d=5a$ range for TMD
two-dimensional materials.  
Non-zero angular momentum $\Delta n$ pairing states can be ground states when the electron and hole densities are unequal. 
The solid red lines mark the points at which adjacent QHI states are equal in energy (the crossing points of QHI lines in Fig.~\ref{fig:E-gap}).
In the strong field limit these lines follow the mid-points of XI state stability range fingers 
that protrude between QHI insulator islands with different Landau level quantum numbers.}
\label{fig:phasediagram}
\end{figure*}

The $ n^\prime -n=m-m^\prime$ selection rule for exchange integrals implies 
coherent electron-hole pair amplitudes can be classified by the difference between
their Landau level indices $\Delta n$.
In the $\Delta n=M$ case, the n-th level in the conduction band 
is coherently coupled only to levels n$^\prime$ in the valence band, such that $n-n^\prime = M$.
At strong magnetic fields, all coherent solutions have a definite value of $\Delta n$, which 
can be interpreted as the exciton angular momentum of the condensed excitons.
The property that solutions with a definite value of $\Delta n$ can be self-consistent is related to 
the rotational symmetry of the models we study.  
In this strong-field regime we will discuss solutions with $\Delta n=0$ and $\Delta n=1$; typical self-consistent 
density matrices for these cases are illustrated in Fig.~\ref{fig:density}.
At weaker fields we find that rotational symmetry can be broken 
by solutions do not conserve $\Delta n$.

The coherence between conduction and valence bands 
develops via spontaneous symmetry breaking,
and serves as a signal (or the order parameter) of excitonic condensation.
All coherent solutions represent different types of excitonic insulator states (XI) 
and compete with incoherent quantum Hall insulator states (QHI) in which both electrons and holes have integer Landau level filling factors. 
Comparisons between total energy per area
\begin{equation}
\label{eq:ground state energy}
\begin{aligned}
        E_0/S = \frac{g}{S}\sum_{n,\tau} &\Big[ \epsilon_{n,\tau} +\frac{(-1)^\tau}{2} \phi_{\tau} \Big] \Big< \rho_{\substack{n n\\ \tau \tau}} \Big>_r\\
        &\ \ + \frac{g}{2S} \sum_{n n^\prime\tau \tau^\prime} U(n,n^\prime;\tau,\tau^\prime) \Big< \rho_{\substack{n n^\prime\\\tau \tau^\prime}} \Big>_r
\end{aligned}
\end{equation}
computed for different self-consistent solutions that start from different seeds identify the mean-field ground state.

The characteristic length and energy scales for quantum particles with
Coulomb interactions are the Bohr radius $a=2\epsilon\hbar^2/e^2m^*$,
and the Rydberg constant $Ry=e^2/2\epsilon a$.
From these we can construct a magnetic field scale $B_0=2\pi\hbar c/ea^2$, 
defined so that there is a quantum of magnetic flux through the area $a^2$.
$B_0$ is the magnetic field scale at which a spatially direct exciton is 
strongly distorted.  If we choose values for $m^*$ and $\epsilon$ that are appropriate for a TMD bilayer encapsulated by
hexagonal boron nitride (hBN) ($m^* \approx$ 0.4 $m_0$, where $m_0$ is the free electron mass, and $\epsilon \approx 5$), we have $a\approx$ 1.3nm, $Ry\approx$ 0.11eV and 
$B_0\approx 2.4\times10^3$T.  These scales are typical for 
TMD semiconductor bilayer devices in experiments.
A similar calculation for III-V semiconductor quantum wells yields $a \approx$ 12nm, $Ry\approx$ 4.6meV, and $B_0 \approx 28$T.

The Coulomb interaction scale ${W_0}/{Ry} = 2(2\pi \,{B}/{B_0})^{\frac{1}{2}}$ 
and the Landau level spacing ${\hbar\omega_c}/{Ry} = 2\pi \,{B}/{B_0}$. 
The ratio between these two quantities increases with decreasing magnetic field.
For example, when $B$=0.1$B_0$, $W_0$=2.52$\hbar \omega_c$; while when $B$=0.01$B_0$, $W_0$=7.98$\hbar \omega_c$. This trend implies that the influence of the discreteness of the Landau level
spectrum gradually increases with increasing magnetic field strength, as we will see 
in the following section, and also reminds us that the lowest Landau level approximation is valid only for sufficiently strong magnetic fields.  
All numerical results are converged with respect to the number of retained Landau.
In our calculations we retain a maximum number of Landau levels
$N_{LL}=15$ and are therefore able to obtain converged results only for 
$B_0/B \lesssim 400$.


\section{Electron-Hole Fluids in the Quantum Hall Regime} \label{sec:result}

\subsection{Integer Total Filling Factors}\label{subsec:int}

In Fig.~\ref{fig:E-gap} we show typical results for the mean-field ground state energies $E_0$ calculated from Eq.~(\ref{eq:ground state energy}),
plotted as a function of the effective gap $\Delta E$ between 
single-particle conduction and valence bands in the 
zero magnetic field limit, when the total charged filling factor $\nu_c$ is a fixed integer.
Since we have in mind mainly the case of quasi-equilibrium spatially indirect exciton condensates in two-dimensional 
TMD bilayers, excitons are first present 
when $\Delta E$ is decreased to equal the
exciton binding energy, and the exciton density increases
if $\Delta E$ decreases further.
Panels (a) and (d) on the left side of 
Fig.~\ref{fig:E-gap} are for the case of layer separation $d=0$; this limit is artificial for spatially indirect condensate since layer isolation,
\textit{i.e.} the absence of interlayer tunneling terms in our model, requires a minimum spatial separation.
We include this limit for comparison purposes only.
Panels (b) and (e) in the middle show the realistic 
layer separation case $d=2a$ at charge neutrality $N_e=N_h$; 
while panels (c) and (f) on the right show results for
the electrostatically doped case $\nu_c$=1.
The top three panels (a-c) are for strong magnetic fields, and the bottom three panels (d-f) are for weaker magnetic fields.

\begin{figure}
\centering
\includegraphics[width=\linewidth]{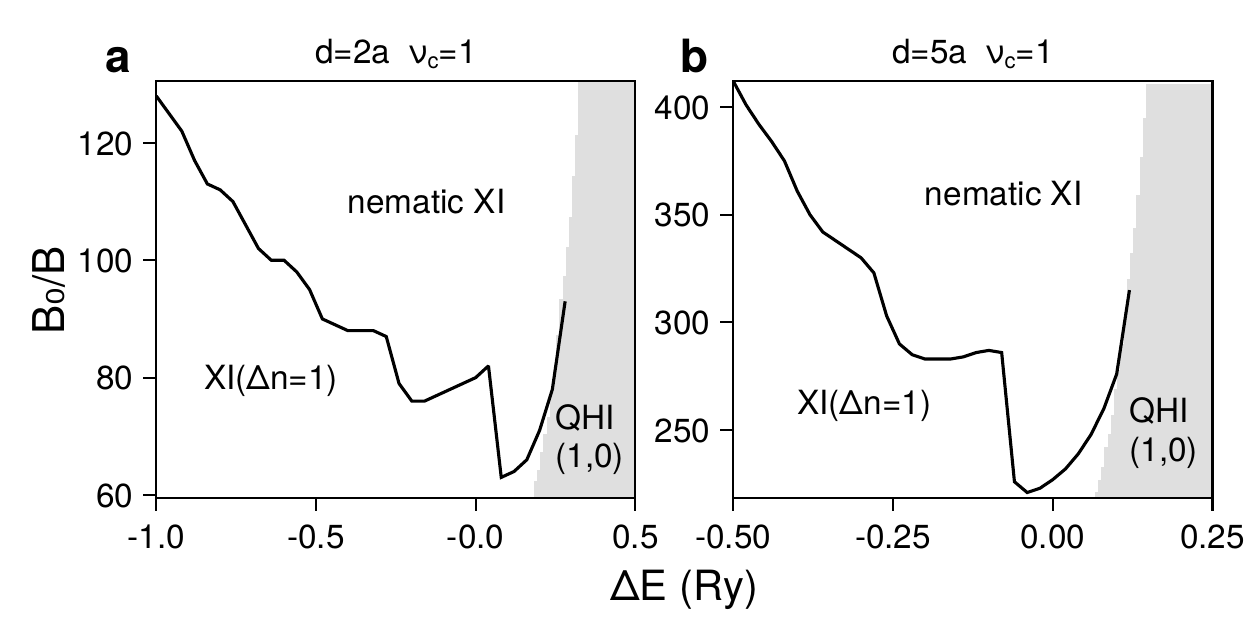}
\caption{Phase diagrams of $\nu_c$=1 case over a weaker magnetic field regime than in Fig.~\ref{fig:phasediagram}.
No QHI states with filling factor ($\nu_e$,$\nu_h$) greater than (1,0) appear in this region of the phase diagram, but we find that 
XI states states that break rotational symmetry (which we refer to as nematic XI states) by coherently mixing condensates 
with different angular momenta are stable over a range of exciton energy that increases with decreasing magnetic field.
The phase boundary (black line) between the $\Delta n$=1 exciton insulator phase and the nematic XI state 
was determined by allowing rotational symmetry to break and noting the properties of the lowest energy 
solution of the mean-field equations.  We do not find nematic XI solutions for $\nu_c=0$.}
\label{fig:weak_field_limit}
\end{figure}

\begin{figure*}
\centering
\includegraphics[width=\linewidth]{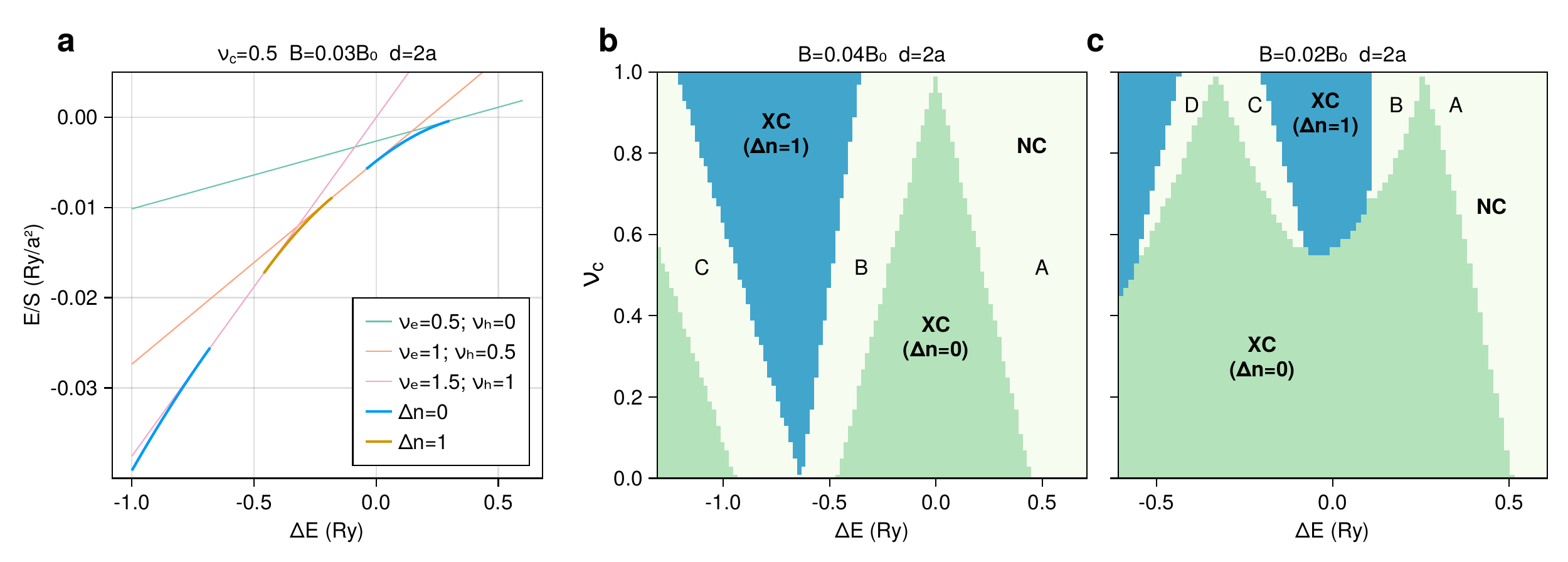}
\caption{(a) Energy per area {\it vs.} gap $\Delta E$ at charge filling factor $\nu_c=0.5$ for $d$=2.0$a$.
(b,c) Phase diagrams {\it vs.} $\nu_c$ and $\Delta E$ at $d$=2.0$a$ and 
$B$=0.04$\,B_0$ and 0.02$\,B_0$, respectively. 
The dark green regions are exciton condensate(XC) phases with pairing angular momentum $\Delta n=0$ and 
the dark blue regions are exciton condensate phases with pairing angular momentum $\Delta n=1$. 
The light green regions are non-condensate (NC) phases distinquished by electron and hole filling factors ($\nu_e,\nu_h$): 
A - ($\nu_c,0$), B - ($1,1-\nu_c$), C - ($1+\nu_c,1$), D - ($2,2-\nu_c$). 
Regions B and C merge at $\nu_c=0$ as the QHI (0,0) phase; regions A and B merge at $\nu_c=1$ as the QHI (1,0) phase.}
\label{fig:non-int filling}
\end{figure*}

The mean-field ground states are either 
quantum Hall insulator (QHI) states with fixed integer filling factors for both electron and hole layers and no interlayer phase coherence,
or excitonic condensates with interlayer coherence (XC), also called excitonic insulators (XI).
We note that the numbers of electrons and holes  
can be read off the slopes of the energy curves 
using 
\begin{equation}
\langle N_{e}+N_{h} \rangle = \left< \frac{\partial H}{\partial \Delta E} \right>= \frac{\partial E_0}{\partial \Delta E}.
\end{equation}
In Fig.~\ref{fig:E-gap}, we can see that $N_e+N_h$ increases monotonically as $\Delta E$ is reduced.  
The energies of the QHI states depend linearly on 
$\Delta E$ as expected, since their carrier densities are independent of $\Delta E$ over their stability ranges.
The coherent excitonic insulator (XI) states appear at 
$\Delta E$ values that are intermediate between those at which 
the QHI states with adjacent integer electron 
and hole filling factors are stable.
The energies of the XI states are not linear in $\Delta E$, because the 
electron and hole filling factors increase from one integer to the next
over their $\Delta E$ stability range. 

As explained in Section ~\ref{sec:method}, 
exciton condensates can be classified 
by $\Delta n$, the integer-valued difference between the Landau level indices of the
electrons and holes that are paired.
The value of $\Delta n$ can be recognized as the angular momentum of the electron-hole pairs.  
The property that the Hartree-Fock solutions have coherence with a fixed $\Delta n$, {\it i.e.,} that there is no mixing between angular momentum pairing channels, follows from the isotropy of our 
two-dimensional electron system model.
At neutrality only $\Delta n=0$ states 
are ground states at any value of layer separation,
although as seen in Fig.~\ref{fig:E-gap}(b,e), $\Delta n \ne 0$ solutions exist as metastable states at some values of layer separation $d$.
Because the $\Delta n = 0$ XI state and $\Delta n = 1$ states have different types of coherence, they connect the $\nu_e=\nu_h=0$ 
QHI state to two different $\nu_e=\nu_h=1$ QHI states, 
shown as two lines with the same slope in the figure. 
At neutrality the $\Delta n=1$ XI solution connects the vacuum state 
to a QHI state that has population inversion and is clearly unstable, {\it i.e.}, a state in which
higher Landau levels are occupied while lower levels are left empty.  
$\Delta n=1$ solutions can, however, be stable when the electron and 
hole filling factors are unequal.  

In Fig.~\ref{fig:phasediagram} we present 
phase diagrams {\it vs.} magnetic field strength and $\Delta E$.
(The phase boundaries in Fig.~\ref{fig:phasediagram} are choppy because the 
set of $\Delta E$ and $B_0/B$ values considered is discrete.)
As the zero-field limit is approached (at the tops of the phase diagrams),
the XI state is stable when $\Delta E$ is 
smaller than the zero-field exciton binding energy and QHI states do not appear.
Because electron and hole densities increase 
with decreasing $\Delta E$, and Landau level degeneracies also increase with increasing $B$, 
a given QHI state's stability range moves to larger $B$ (smaller $B_0/B$) as
$\Delta E$ becomes more negative.  We also see in Fig.~\ref{fig:phasediagram} that
$\Delta n=0$ XI states interpolate between QHI states with the same filling factors for 
electrons and holes, whereas $\Delta n=1$ XI states interpolate between 
$\nu_c=1$ QHI states which have a higher Landau level filling factor for electrons than for holes.


Comparing Fig.~\ref{fig:E-gap}(a-c) with Fig.~\ref{fig:E-gap}(d-f)
we see that the QHI states have smaller stability range in weaker
magnetic fields. 
This property is reflected in the phase diagrams Fig.~\ref{fig:phasediagram} 
through the width of each QHI phase island at fixed $B$, 
which decreases when $B$ is weakened and vanishes at a critical magnetic field strength.
Comparing Fig.~\ref{fig:phasediagram}(a), (b) and 
(c), we see that these critical points move to weaker $B$
as the layer separation increases and exciton binding energies decrease.
Comparing Fig.~\ref{fig:phasediagram}(a) and Fig.~\ref{fig:phasediagram}(d), Fig.~\ref{fig:phasediagram}(b) and Fig.~\ref{fig:phasediagram}(e), or Fig.~\ref{fig:phasediagram}(c) and Fig.~\ref{fig:phasediagram}(f), the 
same decrease in field scale is seen in the transition from $\Delta n=0$ to $\Delta n=1$
phase diagrams, and again signals a reduction in exciton binding energies.

The red lines in Fig.~\ref{fig:phasediagram}
show where adjacent QHI states with adjacent integer filling factors are equal in energy.
At strong fields these lines locate near the centers of the stability regions of the 
XI states that host coherence between the corresponding electron and hole Landau levels.
At weak fields, the QHI states are no longer stable.  The phase diagram is then occupied 
by XI states in which several Landau levels close to the Fermi level participate in coherence,
and the Landau level structure has little impact. 

At weaker fields than those covered in Fig.~\ref{fig:phasediagram} we find that the lowest 
energy solutions of the mean-field equations for $\nu_c=1$ break rotational symmetry by coherently
mixing different values of $\Delta n$.
Fig.~\ref{fig:weak_field_limit} illustrates the competition between non-zero $\Delta n$ and states with mixed $\Delta n$.  Although we have not explicitly verified this property, we believe that
the broken rotational symmetry likely implies that the optimal pairing wavevector 
of these states is non-zero, and that the true ground states states are closely related to 
FFLO states \cite{shimahara1994fulde,fulde1964superconductivity,larkin1965zh,varley2016structure,strashko2020crescent},

\subsection{Non-Integer Filling}\label{subsec:non-int}

We can perform the same mean-field calculations for the case of non-integer charge filling factors $\nu_c$.
Fig.~\ref{fig:non-int filling}(a) summarizes the results obtained when $\nu_c=0.5$.
We note two important differences compared to the integer $\nu_c$ case.  First of all 
insulating states cannot occur at fractional charge filling factor
in mean-field theory.  This property is simply a limitation of mean-field theory since
we do expect that fractional incompressible states 
can be compatible with electron-hole coherence when interactions are treated accurately.  In mean-field theory,
we nevertheless find that coherence is suppressed when the filling factor in either layer 
is close to an integer; freezing one layer or the other by placing a Landau level gap at the Fermi level 
is sufficient to prevent interlayer coherence.  Secondly, solutions with $\Delta n=1$ and 
$\Delta n=0$ appear at different points in the same phase diagram.  

In the quantum Hall regime, mean-field theory is usually reliable at integer filling factors, but 
we know that it is not reliable at fractional filling factors.  In particular, 
we anticipate that fractional QHI states,
not accessible in mean-field-theory calculations, will sometimes \cite{zeng2023evidence} appear between
integer QHI states, especially at larger $d/l$.
Other exotic states \footnote{See for example Ref.\onlinecite{wagner2021s} and work cited therein.}
built from composite-fermion quasiparticles are also a possibility.  
We also anticipate \cite{zou2023skyrmion} that non-uniform density meron and anti-meron 
lattice states that are not captured by the present mean-field calculations, which do not 
allow translational symmetry breaking, will be stable in some regions 
of fractional charge filling factors.

\subsection{The Role of Spin} 

One simplification that we have made, but did not yet comment on, is that we have assumed spinless electrons.
In single layer TMD semiconductors the band extrema are at one of the two 
$K$-points in the triangular lattice Brillouin-zone, where spin-orbit splitting is large.
The two $K$ points define valleys in momentum space that are related to each other 
by time reversal symmetry.  When we refer to spin we mean the 
valley-locked spins of the spin-orbit coupled state. States of opposite spin are therefore in 
opposite valleys.
When the spin degree-of-freedom is included in the absence of a magnetic field, the exciton condensate 
spontaneously spin-polarizes \cite{fernandez1997spin,wu2015theory} when a critical layer separation is exceeded due to 
Coulomb generated exciton-exciton interactions.  Spin-orbit interactions are needed of course if this 
order is to survive at finite temperature.  
In a strong magnetic field, we expect that the tendency toward complete spin-polarization will survive and 
be further supported by the Zeeman coupling effect, which favors particular spin quantum numbers for each type of particle.
According to {\it ab initio} theory the $g$-factors of TMD semiconductor states at the conduction and 
valence band extrema range from $0.5-6$ \cite{deilmann2020ab, wozniak2020exciton, hotger2023spin}, with larger $g$-factors for valence band states.
Once the valence band holes are spin-polarized, they can pair with only one of the two conduction band 
spins and will choose the one with lower energy.  Application of a magnetic field will therefore 
reduce the range of layer separations over which both spins are relevant.  For the strong magnetic fields
of interest here, we expect that exciton condensate states will almost always be fully spin-polarized.
We have tested this expectation by including spin explicitly and performing mean field 
calculations at $B=0.02 B_{0}$ and $d=2a$.  Using realistic valence band $g$-factors, 
we find that the exciton condensate is maximally spin-polarized even if we choose the 
conduction electron $g$-factor to be zero.

\section{Summary and Conclusion}

In the absence of a magnetic field neutral bilayer electron-hole fluids condense at low temperatures 
\cite{zhu1995exciton,naveh1996excitonic, littlewood2004models,perali2013high,combescot2017bose,
zeng2020electrically, ma2021strongly, qi2023thermodynamic} into two-dimensional exciton superfluids provided that the exciton density 
is well below the Mott limit.  In this manuscript we have addressed the fate of exciton condensates 
in the presence of quantizing magnetic fields.  

We have summarized our results in a phase 
diagram, Fig.~\ref{fig:phasediagram}, that specifies where condensation occurs as a function 
of dimensionless field strength $B/B_0$ and reference band gap $\Delta E$.  We choose $\Delta E$ as a 
model parameter instead of exciton density because it is a proxy for exciton chemical potential, 
which can now \cite{ma2021strongly, qi2023thermodynamic} be controlled electrically in two-dimensional materials.
We find that condensation is suppressed in the strong field limit when the electrons and holes 
are able to form low-energy incompressible states (with integer filling of both electron and 
hole Landau levels). These incompressible states appear as a series of stability islands in 
$(B/B_0,\Delta E)$ space distinguished by the integer filling factors of the electron and hole
Landau levels.  The periodic variations between condensed and uncondensed states in Fig.~\ref{fig:phasediagram} are 
the strong-field analogs of the magnetic oscillations discussed in Ref.~\onlinecite{allocca2023fluctuation}.
The phase diagram in Fig.~\ref{fig:phasediagram} is also relevant to the properties of the  
optically pumped electron-hole or polariton fluids \cite{butov2004condensation, deng2002condensation, byrnes2014exciton, xue2016microscopic} if the steady state they form is close to a thermodynamic equilibrium state. 
Fig.~\ref{fig:phasediagram} is constructed using a mean-field theory approach which fails to capture 
the Mott transition to electron-hole fluids at high carrier densities \cite{de2002excitonic, neilson2014excitonic, guerci2019exciton}, 
if that does indeed occur, and 
also fails to capture the special correlations that can  occur within fractionally filled Landau 
levels.  We anticipate that additional stability islands that are not present in our phase diagram
arise in which the incompressible states have fractional Landau level filling factors\cite{zeng2023evidence, wagner2021s}.

It is interesting to compare the exciton condensates studied here with the quantum Hall exciton fluids studied 
experimentally \cite{eisenstein2004bose,eisenstein2014exciton} since the 1990's in electron-electron bilayers.
These have been much more accessible experimentally since they do not require that electrical contact 
be made simultaneously to the conduction band of one two-dimensional semiconductor layer and the 
valence band of a nearby layer.  In the extreme strong field limit in which Landau level mixing can be 
neglected the electron-hole systems studied here and the electron-electron systems studied previously are
identical apart from a difference in the physical meaning of the $\Delta E$ parameter.  In the electron-electron
case, it is universally assumed that both electron layers come to equilibrium with the same electrical reservoir.
The meaning of $\Delta E$ is then simply the difference in external electrical potential between the 
layers.  The meaning of $\Delta E$ is more subtle in our case \cite{zeng2020electrically}, but it is 
equally tunable if the two layers can be contacted to different reservoirs.  It turns out that at 
strong magnetic fields $\Delta E=0$ in the electron case corresponds to the red lines 
in Figs.~\ref{fig:phasediagram}(a) and (b) at the center of the strong field condensate 
stability regions at strong field.  The two cases are most distinct in the weak magnetic field 
regime where the condensate is suppressed in the electron-electron case.  At finite net charge carrier density we find
that the system can break rotational symmetry, as it does in the absence of a magnetic field \cite{strashko2020crescent}.
Broken rotational symmetry is expected \cite{strashko2020crescent} to lead to finite momentum pairing,
which should be explored in future work.

Exciton condensation in separately contacted electron-hole bilayers has been identified experimentally
either by looking for drag effects or for counterflow superfluidity \cite{eisenstein2004bose,eisenstein2014exciton, liu2017quantum, liu2022crossover} in transport,
the technique used most often in bilayer electron systems, or by looking for coherent luminescence at the 
condensate chemical potential, the technique used most often in studies of optically pumped systems.
We expect that the pattern of luminescence to be spatially anisotropic in nematic XI states.
Order enhanced interlayer tunneling, often studied in the electron-electron 
case \cite{spielman2000resonantly, burg2018strongly, lin2022emergence, wang2018electrical}, contributes in the 
electron hole case not to {\em dc} transport, but to  
{\it ac} transport at frequency $\omega$ such that $\hbar \omega$ is equal to the finite exciton chemical potential.  
The strength of the enhanced luminescence is proportional to the overlap of conduction and valence band 
wavefunctions centered in different 2D materials, and will therefore be very sensitive to the number of 
atomic layers of hBN between the electron layer and the hole layer. 

\begin{acknowledgements}
The authors thank Emanuel Tutuc, Kin Fai Mak, and Jie Shan for helpful discussions. 
Work in Austin was supported by the U.S. Department of Energy, Office of Science, Basic Energy Sciences under Award 
DE-SC0021984. The Flatiron Institute is a division of the Simons Foundation.
\end{acknowledgements}

\bibliography{ehfluid.bib}

\end{document}